\documentclass[aps,prl,twocolumn,superscriptaddress,longbibliography]{revtex4-2}

\usepackage{graphicx}%

\usepackage{multirow}
\usepackage{nameref}
\usepackage{siunitx}
\usepackage{amsmath}
\usepackage{xcolor}
\usepackage{natbib}
\usepackage[colorlinks=true,linkcolor=blue,urlcolor=blue,citecolor=blue]{hyperref}

\begin{document}





\title{Single Spin Hall Conversion in the Hall Response of Magnetic-Insulator/Heavy-Metal Bilayers Identified by Odd Spin-Hall-Angle Dependence}


\author{Akashdeep Akashdeep}
\email[]{aakashde@uni-mainz.de}
\affiliation{Institute of Physics, Johannes Gutenberg-University Mainz, 55128 Mainz, Germany}


\author{Ewiese Mohammad Ababneh}
\email[]{eababneh@iu.edu}
\affiliation{Department of Physics, Indiana University, Indianapolis, Indiana 46202, USA}

\author{Botros Gerges Mikhail Fenhas}
\email[]{bfenhas@iu.edu}
\affiliation{Department of Physics, Indiana University, Indianapolis, Indiana 46202, USA}

\author{Duc Minh Tran}
\email[]{tranduc1@uni-mainz.de}
\affiliation{Institute of Physics, Johannes Gutenberg-University Mainz, 55128 Mainz, Germany}

\author{Mathias Kläui}
\email[]{klaeui@uni-mainz.de}
\affiliation{Institute of Physics, Johannes Gutenberg-University Mainz, 55128 Mainz, Germany}
\affiliation{Norwegian University of Science and Technology, 7491 Trondheim, Norway}

\author{Steven S.-L. Zhang}
\email[]{sxz675@case.edu}
\affiliation{Department of Physics, Case Western Reserve University, Cleveland, Ohio, 44106, USA}

\author{Vivek Amin}
\email[]{vpamin@iu.edu}
\affiliation{Department of Physics, Indiana University, Indianapolis, Indiana 46202, USA}

\author{Gerhard Jakob}
\email[]{jakob@uni-mainz.de}
\affiliation{Institute of Physics, Johannes Gutenberg-University Mainz, 55128 Mainz, Germany}

\author{Timo Kuschel}
\email[]{tkuschel@uni-mainz.de}
\affiliation{Institute of Physics, Johannes Gutenberg-University Mainz, 55128 Mainz, Germany}



\begin{abstract}

The Hall response of magnetic-insulator/heavy-metal bilayers is usually ascribed to two successive spin Hall conversions and is therefore expected to be even in the spin Hall angle. Using distinct magnetic-field windows, we separate the Hall voltage of yttrium iron garnet bilayers with Ta and Pt of opposite spin Hall angle into contributions from  magnetization- and magnetic-field-driven precession as well as from ordinary Hall effect. The field-driven component reverses sign between Ta and Pt, revealing a Hall response odd-in-spin-Hall-angle. The magnetization-driven component also has opposite signs for Ta and Pt, except for the thinnest heavy-metal layers. A drift-diffusion model with interface-generated spin currents describes this behavior: the field-driven component is predominantly carried by a single spin Hall conversion, and the magnetization-driven component is consistent with one as well. Other potential explanations, based on a description with a finite imaginary spin-mixing conductance rather than interface-generated spin currents, require unusually large values of this conductance, making this mechanism unlikely. Our findings advance the understanding of Hall responses in magnetic-insulator/heavy-metal bilayers toward a more general description.

\end{abstract}


\maketitle


\textbf{\textit{Introduction}}---Magnetic-insulator\slash heavy-metal (MI/HM) bilayers combine the magnetic order parameter of an electrically insulating magnet with efficient spin-to-charge conversion in an adjacent HM. Because the MI carries no charge current, as is the case for the prototypical yttrium iron garnet ($\text{Y}_3\text{Fe}_5\text{O}_{12}$, YIG), transverse Hall signals in MI/HM bilayers are a particularly clean electrical probe of interfacial spin transport, magnetic order, and spin-dependent scattering \cite{PhysRevB.87.144411, PhysRevLett.110.206601, PhysRevB.87.224401, PhysRevLett.99.126601}. They enable non-invasive magnetic-state readout, interfacial exchange diagnostics, spin--orbit-torque metrology, and sensing in MI heterostructures \cite{doi:10.1021/acs.nanolett.8b05121, Ngaloy2025, Shiino2024, 10.1063/5.0049235, PhysRevApplied.17.064038, PhysRevB.104.224410, 10.1063/1.4916342, 10.1063/5.0132895, https://doi.org/10.1002/advs.202403852, s23p-y6fd, PhysRevB.94.134418, PhysRevB.94.094401, PhysRevApplied.6.034007, PhysRevB.103.134406}. In HMs such as Pt and Ta, strong spin--orbit coupling interconverts charge currents, spin currents, and spin accumulations through the spin Hall effect (SHE) and inverse spin Hall effect (ISHE) \cite{PhysRevLett.83.1834, 10.1063/1.2199473, valenzuela2006direct, 6516040, RevModPhys.87.1213, PhysRevLett.94.047204, doi:10.1126/science.1218197, doi:10.1126/science.1105514}, with an efficiency set by the spin Hall angle $\theta_{\text{sh}}$, which depends on the HM, its conductivity, disorder, crystallinity, and thin-film phase \cite{PhysRevB.87.224401, PhysRevLett.100.096401, PhysRevB.98.060410, PhysRevB.94.060412, PhysRevB.111.104421}.

These Hall signals are conventionally interpreted through the spin Hall magnetoresistance (SMR) and its underlying double spin Hall mechanism. A longitudinal charge current in the HM first generates a transverse spin current and spin accumulation by the SHE. The accumulation then interacts with the magnetic order at the MI/HM interface through spin reflection, absorption, dephasing, spin-flip scattering, and the complex spin-mixing conductance $G_{\uparrow\downarrow}=G_r+iG_i$ \cite{PhysRevB.86.214424, PhysRevLett.111.176601, PhysRevB.87.144411, PhysRevLett.110.206601, PhysRevB.87.224401}. The remaining spin current is converted back into a transverse charge signal by the ISHE \cite{PhysRevB.87.224401, PhysRevB.87.144411, doi:10.1021/acs.nanolett.0c02834, PhysRevB.87.174417, 10.1063/1.4813760}. Because two spin Hall conversions are involved, the Hall response is expected to be even in $\theta_{\text{sh}}$, scaling as $\theta_{\text{sh}}^{2}$.

Within this picture, the Hall response associated with an out-of-plane magnetization component has been discussed as the SMR-based spin Hall anomalous Hall effect (SH-AHE) \cite{PhysRevB.87.224401, PhysRevB.87.144411, 10.1063/1.4813760, 10.1063/1.4916342}. The nonlinear response from Hanle-type spin precession around the applied magnetic field $\boldsymbol{B}_0$ has been termed spin Hall Hanle effect (SH-HE) \cite{PhysRevLett.116.016603, PhysRevB.94.174407, PhysRevB.106.184420}. Both add to the Lorentz-force-driven ordinary Hall effect (OHE), which provides a linear background \cite{PhysRevB.87.174417}. Earlier Hall data on YIG/Pt and YIG/Ta already show material- and temperature-dependent signs and a contribution that continues to evolve far above the saturation field of the YIG magnetization \cite{PhysRevB.89.140407, PhysRevB.94.174430}, but the parity of these signals in $\theta_{\text{sh}}$ and the origin of the nonsaturating part have remained unresolved.

In this Letter, we report a Hall response in MI/HM bilayers with an odd $\theta_{\text{sh}}$ dependence, which challenges the established double-SHE picture. We describe it within a theoretical framework that includes interface-generated spin currents (IGSCs), which can produce an odd $\theta_{\text{sh}}$ dependence through a single SHE process \cite{PhysRevLett.116.136601, PhysRevLett.121.136805, doi:10.1021/acs.nanolett.0c02834, PhysRevB.94.140413}. The model captures the main features of the measured Hall responses and thereby extends the SMR picture toward a description encompassing both odd and even $\theta_{\text{sh}}$ dependencies. This broader understanding of interfacial magnetotransport will help to develop new functionalities and material combinations for spintronic devices based on MI/HM bilayers.

\begin{figure*}
    \centering
    \includegraphics[width=\textwidth]{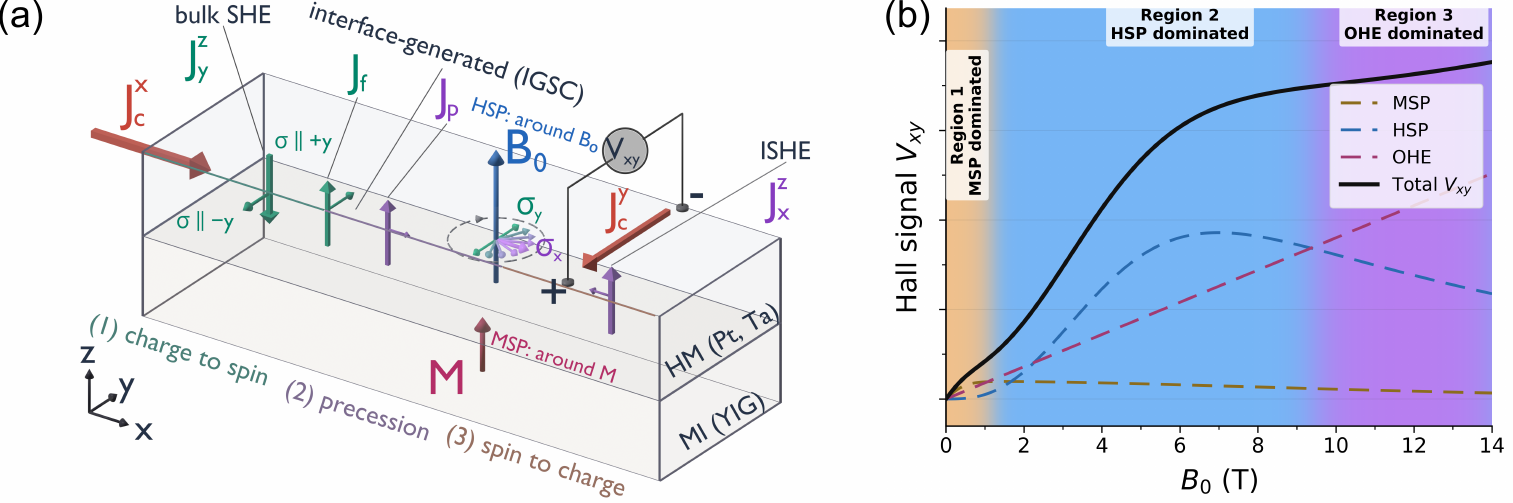}
    \caption{Classification of Hall responses in MI/HM bilayers.
    (a) Schematic of charge-to-spin (left), spin precession (center), and spin-to-charge conversion (right) in MI/HM bilayers. Green arrows denote spin currents polarized along $\hat{y}$ (bulk SHE, $J_y^z$, and the interfacial filtering current $J_{\text{f}}$); purple arrows denote spin currents polarized along $\hat{x}$ (the interfacial precession current $J_{\text{p}}$). Only the $\hat{x}$-polarized component is converted by the ISHE into the measured transverse charge current $J_c^y$ and Hall voltage $V_{xy}$.
    (b) Schematic Hall response as a function of the external magnetic field $B_0$, illustrating how MSP-, HSP-, and OHE-dominated contributions can be approximately separated using characteristic magnetic-field windows.}
    \label{fig:1}
\end{figure*}

\textbf{\textit{Three fundamental categories of transport effects}}---We group the relevant Hall responses into three categories: magnetization-dependent spin precession (MSP), external-magnetic-field-dependent spin precession (HSP), and the Lorentz-force contribution as the OHE. In MSP, charge-to-spin conversion is followed by spin precession governed primarily by the magnetization direction, after which spin-to-charge conversion produces a transverse signal. In HSP, the spin accumulation precesses under $\boldsymbol{B}_0$, including Larmor/Hanle precession, before conversion \cite{PhysRevLett.116.016603, PhysRevLett.99.126601, PhysRevB.94.174407, PhysRevB.106.184420}. In both, the conversion steps may occur in the HM bulk via SHE and ISHE, or at the MI/HM interface via interface-generated spin-dependent scattering (IGSS) or IGSCs [Fig.~\ref{fig:1}(a)].

MSP and HSP are therefore phenomenological categories rather than unique signatures of a specific microscopic mechanism. The conventional SHE+ISHE scenario produces an even-in-$\theta_{\text{sh}}$ response because both conversion steps are spin Hall processes, whereas interface-related mechanisms---IGSS, or an IGSC followed by ISHE conversion---can result in an odd dependence on $\theta_{\text{sh}}$ \cite{PhysRevLett.116.136601, PhysRevLett.121.136805, PhysRevB.87.144411, doi:10.1021/acs.nanolett.0c02834, PhysRevB.94.140413, 4m5d-ylyr}. We therefore compare SHE+ISHE with IGSC+ISHE as a representative test of alternative microscopic origins and separate the Hall response into three magnetic-field regimes [Fig.~\ref{fig:1}(b)]: a low-field MSP-dominated regime up to the out-of-plane saturation field of the MI, an intermediate-field HSP-dominated regime, and a high-field OHE regime providing the linear background.

\textbf{\textit{Methods}}---We investigate this ambiguity using bilayers of YIG with Ta or Pt, which provide a sign-sensitive test because the two HMs have opposite $\theta_{\text{sh}}$ while sharing the same MI platform \cite{PhysRevB.87.174417, PhysRevB.98.060410, PhysRevB.94.060412, PhysRevB.90.174436}. The heterostructures consist of YIG(\SI{25}{\nano\meter})/HM on non-magnetic insulating $\text{Gd}_3\text{Ga}_5\text{O}_{12}$ (GGG)(001) substrates, with $t_{\text{HM}} = \SIrange{4}{12}{\nano\meter}$. Sample growth and structural characterization are given in Sec.~A, the longitudinal resistivity and $\beta$-Ta phase assignment in Sec.~B, and angular-dependent magnetoresistance scans confirming an SMR response that excludes a dominant magnetic proximity effect at room temperature in Sec.~C of the Supplemental Material (SM)~\cite{SM}, which includes Refs.~\cite{PhysRevApplied.10.024051, PhysRevB.97.020403, WEN2025173333, 10.1063/1.4773509, PhysRevLett.115.097401, 7372460, PhysRevB.93.214415, 10.1063/1.4987145, PhysRevLett.119.227205, PhysRevB.102.214438}.

\begin{figure*}
    \centering
    \includegraphics[width=\textwidth]{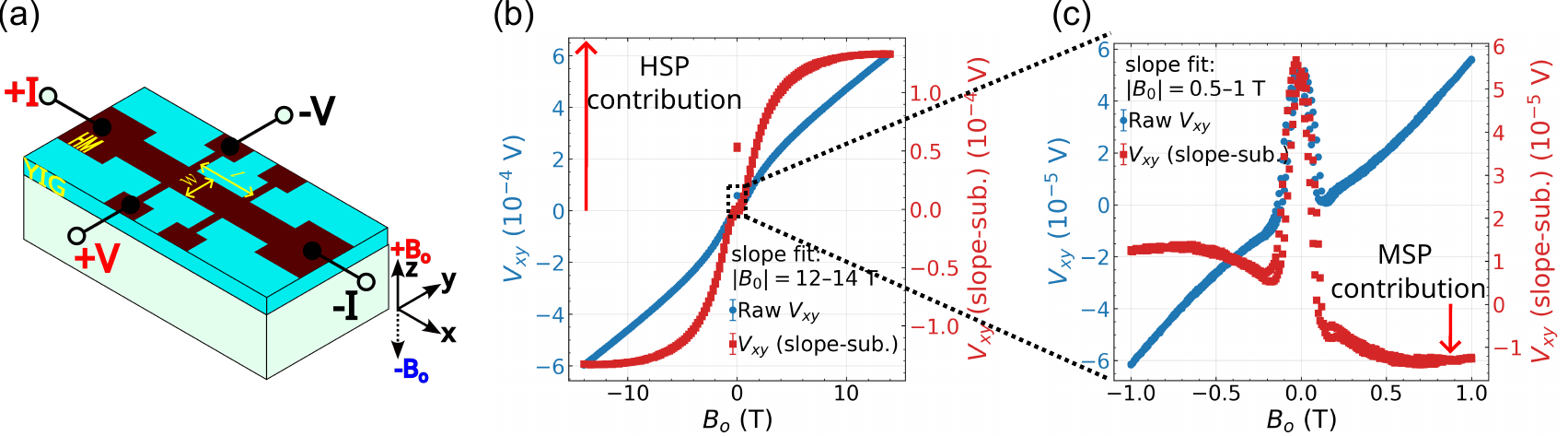}
    \caption{Measurement geometry and representative Hall data.
    (a) Hall-bar geometry showing the longitudinal current path and transverse voltage probes under an out-of-plane magnetic field $\boldsymbol{B}_0$.
    (b,c) Hall measurements of YIG/Ta(\SI{4}{\nano\meter}): (b) Hall voltage $V_{xy}$ vs.\ magnetic field up to $\pm$14~T: raw data (blue) and nonlinear remaining data (red) after subtraction of the linear slope fitted for $|B_0| > \SI{12}{\tesla}$, highlighting the HSP contribution. (c) $V_{xy}$ vs.\ magnetic field up to $\pm$1~T: raw data (blue) and the remainder (red) after subtraction of the linear background fitted for $\SI{0.5}{\tesla} < B_0 < \SI{1}{\tesla}$, whose antisymmetric (broad step-like) part is the MSP contribution. The even-in-$B_0$ peak at $|B_0| < \SI{0.25}{\tesla}$ arises from transverse SMR, which depends on the product $m_x m_y$ of the in-plane YIG magnetization components.}
    \label{fig:1b}
\end{figure*}

Magnetotransport measurements were performed at room temperature in a $\pm\SI{14}{\tesla}$ superconducting magnet cryostat using a right-handed coordinate system ($I \parallel \hat{x}$, $V_{xy} \parallel \hat{y}$, $\boldsymbol{B_0} = B_0\hat{z}$) with the Hall-bar geometry of Fig.~\ref{fig:1b}(a). A constant current $I = \SI{1}{\milli\ampere}$ ($J_{\text{max}} = \SI{1.25e10}{\ampere\per\meter\squared}$) was applied, and $V_{xx}$ and $V_{xy}$ were acquired in delta mode to suppress thermoelectric artifacts. A first separation of the MSP, HSP, and OHE responses uses the field windows of Fig.~\ref{fig:1}(b) \cite{PhysRevLett.116.016603, PhysRevB.94.174407}. The intermediate-field HSP-dominated signal exhibits a systematic sign reversal between YIG/Ta and YIG/Pt, indicating an odd dependence on $\theta_{\text{sh}}$, while the low-field MSP-dominated response shows a more complex thickness dependence in which thicker HM films also reveal the sign change between YIG/Ta and YIG/Pt. These results discussed below challenge the uniqueness of the double-spin-Hall interpretation.

\textbf{\textit{Results}}---Figure~\ref{fig:1b}(b) shows the Hall voltage $V_{xy}$ of YIG/Ta(\SI{4}{\nano\meter}) up to $\pm$\SI{14}{\tesla}. The raw data (blue dots) contain the OHE, the MSP signal, and an additional nonlinear contribution from the HSP signal. To isolate this part, the high-field slope is determined from $|B_0| > \SI{12}{\tesla}$, for which the magnetization is saturated, and subtracted from the full Hall curve; the resulting nonlinear component (red squares) reveals the HSP signature \cite{PhysRevLett.116.016603, PhysRevB.94.174407, PhysRevB.106.184420}. Because the HSP contribution peaks below or near \SI{12}{\tesla}, this high-field window largely separates the OHE from the HSP. The MSP contribution, in contrast, saturates once the YIG magnetization is fully aligned out of plane, i.e., above the out-of-plane saturation field ($\approx\SI{0.25}{\tesla}$); the nonlinear contribution that we assign to HSP therefore develops at magnetic field strengths far larger than needed to saturate the magnetization and cannot originate from the MSP.

The close-up in Fig.~\ref{fig:1b}(c) shows the low-field Hall voltage up to $\pm$\SI{1}{\tesla}. Here the raw data (blue dots) contain a linear background, dominated by OHE and HSP, which is obtained by fitting between \SI{0.5}{\tesla} and \SI{1}{\tesla}, in which range the magnetization-dependent contribution is saturated, and then subtracted. The antisymmetric part of the remainder (red squares) is the MSP signal; the symmetric peak is the transverse SMR associated with the in-plane YIG magnetization \cite{10.1063/1.4813760, PhysRevB.87.184421}.


\textbf{\textit{Theory}}---To describe the full Hall response as a combination of HSP, MSP, and OHE contributions, we model the bilayer with a semi-classical drift-diffusion formalism in an external magnetic field, treating the system as quasi-one-dimensional ($x$ and $y$ semi-infinite, $z$ finite). As sketched in Fig.~\ref{fig:1}(a), an electric field along $\hat{x}$ generates spin currents flowing along $\hat{z}$, via the bulk SHE and the IGSCs; their spin accumulation undergoes Larmor precession and is reconverted into a transverse charge current by the ISHE. The drift-diffusion equation for the spin accumulation reads
\begin{equation}
      -\frac{\partial^2 \mu_{i}(z)}{\partial z^2}+\frac{\mu_i(z)}{l^2_s}+\frac{(\Omega \times \mu)_i}{D}=0,
      \label{dif}
\end{equation}
with the spin accumulation $\mu_{i}(z)$ (in Joules) polarized along $\hat{i}$, the spin diffusion length $l_s$, the diffusion constant $D$ of the HM, and the Larmor precession frequency $\Omega$. The spin current density along $\hat{z}$ and the charge current density are
\begin{equation}
      J_{i}^{z}(z)=-\frac{\sigma}{2e}\frac{d\mu_{i}(z)}{dz}+J_{i}\delta_{iy},
      \label{se}
\end{equation}
\begin{equation}
      J_{c}^{j}=\sigma E_j\delta_{jx}+\frac{\theta_{\text{sh}}\sigma}{2e} (\nabla \times \mu)_j ,
      \label{cu}
\end{equation}
with the charge conductivity $\sigma$, the elementary charge $e$, and the Kronecker delta $\delta$. The bulk SHE spin current density is $J_{i}=\theta_{\text{sh}}\sigma E_{x}$, with the applied electric field $E_x$. Superscripts denote the flow direction and subscripts the spin-polarization direction ($i,j\in\{x,y,z\}$) or the charge $c$.

The spin current vanishes at the HM/vacuum boundary, $J^z_i(0)=0$. At the HM/MI interface, we apply magnetoelectronic circuit theory, for which interfacial spin transport is governed by the spin-mixing conductance $G_{\uparrow\downarrow}=G_r+iG_i$, and we add the IGSCs: a spin-filtering current $J_{\text{f}}$ (spin-$y$, flowing along $\hat{z}$) and a spin-precession current $J_{\text{p}}$ (spin-$x$, flowing along $\hat{z}$), giving at $z=t$
\begin{align}
    \begin{bmatrix}
       J^z_x(t)\\J^z_{y}(t)
 \end{bmatrix}
    =
\frac{1}{e}\begin{bmatrix}
      G_r&&-G_i\\
       G_i&&G_r
 \end{bmatrix}
 \begin{bmatrix}
     \mu_x\\\mu_y
 \end{bmatrix}
   +
   \begin{bmatrix}
       J_{\text{p}}\\J_{\text{f}}
\end{bmatrix}\label{eq7}
\end{align}
with $J_{\text{p,f}}=\theta_{\text{p,f}}\sigma E_x$ and $\theta_{\text{p,f}}$ dimensionless. Without a magnetic field, symmetry forbids a voltage along $y$: a spin-$x$ current flowing along $\hat{z}$ is required, provided here by precession around either $\boldsymbol{B}_0$ or the interfacial magnetization $\boldsymbol{M}$ [Fig.~\ref{fig:1}(a)], the latter governed by $G_i$ and/or $J_{\text{p}}$. The Hall voltage $V_{xy}=(w/\sigma)J_c^y(t)+V_{\text{OHE}}$, with the HM width $w$ and $J_c^y(t)=(1/t)\int_0^t (\theta_{\text{sh}}\sigma/2e)(d\mu_x/dz)\,dz$, then separates into

\begin{equation}
        V_{xy}=V_{\text{HSP}}+V_{\text{MSP}}+V_{\text{OHE}}.
    \label{eq:sum}
\end{equation}
Here, $V_{\text{HSP}}$ arises from precession around $\boldsymbol{B}_0$, $V_{\text{MSP}}$ from precession around the magnetization $\boldsymbol{M}$, and $V_{\text{OHE}}=\sigma E_x w R_{H} B_0$ is the ordinary Hall voltage with the ordinary Hall coefficient $R_H$.

Solving Eqs.~(\ref{dif})--(\ref{eq7}) shows that both precession voltages split into a term linear in $\theta_{\text{sh}}$ and a term quadratic in $\theta_{\text{sh}}$. The linear terms are
\begin{equation}
V_{\text{HSP}}^{\text{l}}=- \theta_{\text{sh}} \theta_\text{f}  \Lambda \sigma\sinh^2\!\left(\frac{ \Lambda t}{2}\right),
\label{eq:odd_HSP}
\end{equation}
\begin{equation}
    V_{\text{MSP}}^{\text{l}}=- \theta_{\text{sh}}\sinh^2\!\left(\frac{ \Lambda t}{2}\right) \bigl[ \theta_\text{p}\mathcal{D}_0 + G_i \theta_\text{f}\cosh( \Lambda t) \bigr],
    \label{eq:odd_MSP}
\end{equation}
with $\mathcal{D}_0 = G_r \cosh(\Lambda t) + (\Lambda \sigma/2) \sinh(\Lambda t)$ and the scaled inverse spin-diffusion length $\Lambda(B_0) = l_s^{-1}(1+i\,eB_0 l^2_s/m_e D)^{1/2}$. They are carried by the IGSCs ($\theta_{\text{f}}$, $\theta_{\text{p}}$), i.e., a single SHE conversion occurs via the ISHE. The quadratic counterparts, $V_{\text{HSP}}^{\text{q}} \propto \theta_{\text{sh}}^{2}$ (with both a $G_r$-independent and a $G_r$-proportional term) and $V_{\text{MSP}}^{\text{q}} \propto \theta_{\text{sh}}^{2} G_i$, correspond to the conventional double SHE conversion; the full expressions and derivation are given in Sec.~D of the SM. The linear terms therefore produce an odd $\theta_{\text{sh}}$ dependence, provided that the sign of $\theta_{\text{p,f}}$ is the same for Ta and Pt.

\begin{figure}
    \centering
    \includegraphics[width=\columnwidth]{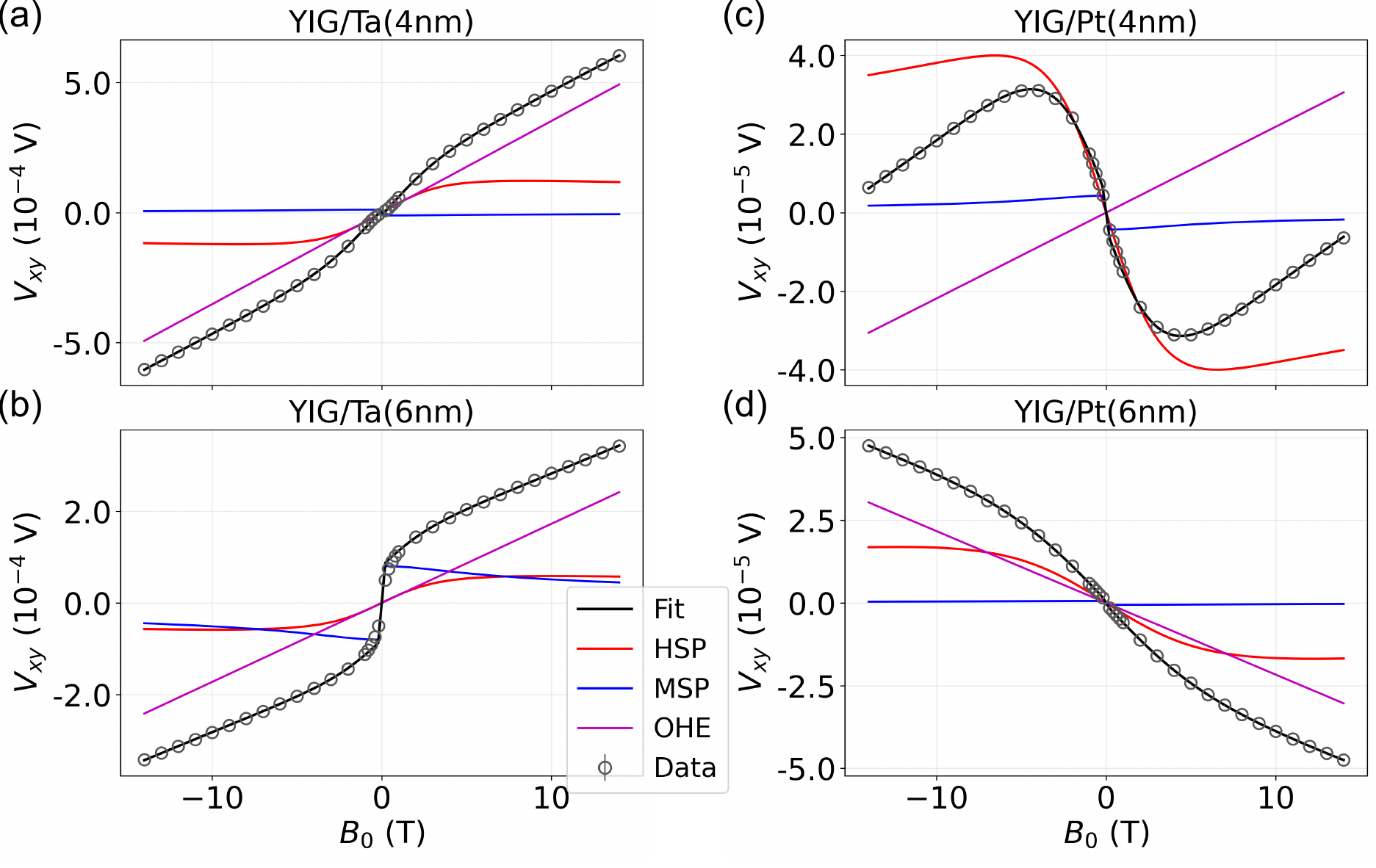}
    \caption{Decomposition of the field-antisymmetric Hall voltage $V_{xy}$ for YIG/HM bilayers, measured over $\pm\SI{14}{\tesla}$. Gray open circles are the measured data (all data points for $\lvert B_0 \rvert < \SI{1}{\tesla}$, one point per tesla for $\lvert B_0 \rvert > \SI{1}{\tesla}$; error bars are smaller than the symbols), and the black line is the total fit with $G_i = 0$ (Table~\ref{tab:1}), decomposed into the HSP ($V_{\mathrm{HSP}}$, red), MSP ($V_{\mathrm{MSP}}$, blue), and OHE ($V_{\mathrm{OHE}}$, magenta) contributions. (a,b) YIG/Ta and (c,d) YIG/Pt, for HM thicknesses of (a,c) \SI{4}{\nano\meter} and (b,d) \SI{6}{\nano\meter}.}
    \label{fig:3}
\end{figure}

\textbf{\textit{Data fitting}}---Figure~\ref{fig:3} shows the fitted decomposition of $V_{xy}$. $V_{\text{OHE}}$ (magenta) changes sign in Pt between \SI{4}{\nano\meter} and thicker films. $V_{\text{HSP}}$ (red) flips sign between Pt and Ta, which the fit attributes to the opposite sign of $\theta_{\text{sh}}$ in the linear term [Eq.~(\ref{eq:odd_HSP})], i.e., a single SHE mechanism. $V_{\text{MSP}}$ (blue) resembles a step function at low magnetic fields because it originates from precession around the magnetization rather than the magnetic field, and decays at high magnetic fields due to spin dephasing.

We fitted each thickness series globally to the part of $V_{xy}$ odd in $B_0$; the complete series and component breakdowns are given in Secs.~E and F of the SM. The spin Hall conductivity $\theta_{\text{sh}}\sigma$ was fixed for each metal with $\theta_{\text{sh}}(\SI{4}{\nano\meter}) = 0.042$ for Pt \cite{PhysRevB.91.115316} 
and $-0.020$ for Ta \cite{PhysRevB.87.174417}, and $G_r$ was fixed to literature values, $0.62\times10^{14}\,\Omega^{-1}\mathrm{m}^{-2}$ (Pt) and $1.53\times10^{14}\,\Omega^{-1}\mathrm{m}^{-2}$ (Ta) \cite{PhysRevB.87.174417}. For each sample, we fitted $\theta_{\text{f}}$, the spin relaxation time $\tau_s = l_s^2/D$, $R_H$, and either $G_i$ or $\theta_{\text{p}}$, with $l_s \propto 1/\rho$ and one prefactor per metal. Both variants describe the data with similar quality, but the $G_i$ variant requires large $G_i/G_r$ (see below), so we present the variant with $\theta_{\text{p}}$, the interface-generated spin-orbit precession current, and $G_i = 0$ (Table~\ref{tab:1}; both variants in Table~SI and Figs.~S5 and S6 of the SM). The fitted $\theta_{\text{p}}$ keeps a consistent sign for Pt and Ta across the thickness series, except for Ta at \SI{4}{\nano\meter}, and $\theta_{\text{f}}$ stays positive for both metals without reaching its lower bound, despite their opposite $\theta_{\text{sh}}$.

\begin{table*}[t]
\begin{ruledtabular}
\begin{tabular}{cccccc}
$t$ (nm) & $l_s$ (nm) & $\theta_{\text{f}}$ & $\theta_{\text{p}}$ ($10^{-2}$) & $D$ ($10^{-6}\,\mathrm{m^2/s}$) & $R_H$ ($10^{-10}\,\mathrm{m^3/C}$) \\
\hline
\multicolumn{6}{l}{YIG/Pt, $G_r = 0.62\times10^{14}\,\Omega^{-1}\mathrm{m}^{-2}$ (fixed)} \\
4  & $0.820\pm0.029$ & $0.220\pm0.026$ & $0.517\pm0.071$ & $0.422\pm0.051$ & $\phantom{-}0.0873\pm0.0043$ \\
6  & $1.027\pm0.037$ & $0.229\pm0.031$ & $0.195\pm0.030$ & $1.25\pm0.14$ & $-0.1301\pm0.0024$ \\
8  & $1.163\pm0.042$ & $0.226\pm0.032$ & $0.172\pm0.030$ & $1.54\pm0.17$ & $-0.1570\pm0.0019$ \\
12 & $1.277\pm0.046$ & $0.220\pm0.030$ & $0.134\pm0.058$ & $1.45\pm0.18$ & $-0.1789\pm0.0012$ \\
\hline
\multicolumn{6}{l}{YIG/Ta, $G_r = 1.53\times10^{14}\,\Omega^{-1}\mathrm{m}^{-2}$ (fixed)} \\
4  & $0.79\pm0.26$ & $0.169\pm0.068$ & $-0.66\pm0.18$ & $0.32\pm0.13$ & $1.407\pm0.045$ \\
6  & $0.89\pm0.30$ & $0.227\pm0.089$ & $\phantom{-}11.8\pm2.1$ & $0.44\pm0.20$ & $1.035\pm0.039$ \\
8  & $1.04\pm0.35$ & $0.29\pm0.11$ & $\phantom{-}14.1\pm2.8$ & $0.70\pm0.30$ & $0.662\pm0.032$ \\
12 & $1.09\pm0.37$ & $0.34\pm0.12$ & $\phantom{-}12.6\pm2.6$ & $0.66\pm0.27$ & $0.634\pm0.022$ \\
\end{tabular}
\caption{Transport parameters for Pt and Ta from the fits with $G_i = 0$: spin diffusion length $l_s$, strengths $\theta_{\mathrm{f}}$ and $\theta_{\mathrm{p}}$ of the interface-generated spin filter and precession currents, diffusion constant $D = l_s^2/\tau_s$ derived from the fitted spin relaxation time $\tau_s$, and ordinary Hall coefficient $R_H$. Uncertainties include systematic shifts from the $G_r$ and $\rho$ uncertainties (Sec.~G of the SM); Table~SI of the SM lists $\tau_s$ and compares both fit variants.\label{tab:1}}
\end{ruledtabular}
\end{table*}

\textbf{\textit{Discussion}}---The ordinary Hall coefficient $R_H$ and its sign agree with the literature. For Ta, $R_H$ remains positive across all thicknesses, reflecting dominant hole carriers at the Fermi surface \cite{bahrami2025vortexmotioninducedlosses, osti_6303727, FermiSurfaceTa, szurek2025roleelectroninteractionsfailed, 10.1116/1.569114}. In Pt, light s-like electron and heavier d-like hole states compete \cite{FermiSurfacePt}, so that thicker films are electron-like with negative $R_H$, whereas in the thin-film limit interface and grain-boundary scattering can favor either sign \cite{PhysRevB.2.883, 10.1063/1.1657635, 10.1063/1.4916342, dushenko2018tunable, PhysRevB.22.6065}. Accordingly, $R_H$ is positive in our thinnest Pt film and negative in the thicker ones (Table~\ref{tab:1}). Determining $R_H$ for each sample is therefore essential, as it fixes the sign convention against which the $\theta_{\text{sh}}$ dependence of the HSP and MSP contributions is assessed (Figs.~\ref{fig:1b}--\ref{fig:3}).

Our results are inconsistent with a purely $\theta_{\text{sh}}^2$ (double-SHE) framework, suggesting the presence of single-SHE contributions. In principle, a finite $G_i$ could also produce MSP-like Hall signals through interfacial exchange, affecting the observed sign and thickness trends. However, fitting a nonzero $G_i$ (with $\theta_{\text{p}} = 0$) at the fixed $G_r$ yields $|G_i|/G_r \approx 0.09$--$0.8$ across the series, well above the values reported for HM/YIG interfaces, where the imaginary component is typically very small ($G_i/G_r \approx 0.025$--$0.065$, often approximated as $G_i \approx 0$) \cite{Jia_2011, 10.1063/1.5053902, 10.1063/1.4813760, PhysRevB.87.224401, 10.1063/1.4916342} (Table~SI and Fig.~S6 of the SM). By contrast, a finite $\theta_{\text{p}}$ (with $G_i = 0$), a precessing IGSC acting as an additional spin-current source, reproduces the same Hall-voltage trends with $G_r$ fixed to literature values. Here, $\theta_{\text{p}}/|\theta_{\text{sh}}| \approx 0.05$--$0.12$ for Pt, whereas $\theta_{\text{p}}/|\theta_{\text{sh}}| \approx 6.7$--$9.3$ for Ta (6--\SI{12}{\nano\meter}). The larger $\theta_{\text{p}}$ in Ta reflects a measured MSP step that is itself about two orders of magnitude larger than in Pt. Because $\theta_{\text{f}}$ is comparable in both metals, whereas $J_{\text{p}}$, unlike $J_{\text{f}}$, requires a magnetic layer \cite{PhysRevLett.121.136805}, this contrast may reflect a stronger coupling to the YIG magnetization at the Ta interface, e.g., through interfacial exchange or interface chemistry \cite{Liu_2018, Yuasa_2018, PhysRevB.87.174417}, rather than the spin--orbit strength alone; its microscopic origin remains open.

The thickness-averaged spin diffusion lengths, $\overline{l_s}(\mathrm{Ta}) = \SI{0.96}{\nano\meter} \pm \SI{0.32}{\nano\meter}$ and $\overline{l_s}(\mathrm{Pt}) = \SI{1.07}{\nano\meter} \pm \SI{0.04}{\nano\meter}$, agree with reported values. The Ta value lies at the lower end of those reported for the highly resistive $\beta$-Ta phase \cite{PhysRevB.91.144412, PhysRevB.87.174417, PhysRevB.94.054416, PhysRevB.98.060410}, and the Pt value lies in the range expected when interface spin loss or high resistivity limits spin diffusion \cite{10.1063/1.4848102, PhysRevB.91.024402, PhysRevB.94.060412}. The assumed scaling $l_s \propto 1/\rho$ is expected for Elliott--Yafet spin relaxation \cite{PhysRevB.94.060412, PhysRevLett.116.126601}. 

The HSP and MSP contributions have opposite signs in Pt and Ta, i.e., they are odd in $\theta_{\text{sh}}$, except for the MSP of Ta at \SI{4}{\nano\meter}, which the fit captures as a sign reversal of $\theta_{\text{p}}$ and which may reflect a competition between $\theta_{\text{p}}$ and a small finite $G_i$ at the lowest thickness. Opposite room-temperature signs of the low-field anomalous-Hall-like signal were previously reported for YIG/Pt(\SI{15}{\nano\meter}) and YIG/Ta(\SI{6}{\nano\meter}) \cite{PhysRevB.89.140407}, independently consistent with our MSP signs above \SI{4}{\nano\meter}; to our knowledge, a systematic Ta-thickness-dependent sign reversal of the MSP has not been established before, and its microscopic origin remains open. In the HSP, the linear term exceeds the quadratic one in magnitude for all films; in Pt the two terms have opposite signs, so the measured HSP sign is set by the linear term, with $\theta_{\text{f}}$ positive for both metals (0.17--0.34). Taken together, these results favor a single-SHE mechanism carried by the IGSCs, $\theta_{\text{f}}$ in the HSP and $\theta_{\text{p}}$ in the MSP, which reproduces the odd-in-$\theta_{\text{sh}}$ Hall response without a finite $G_i$.

\textbf{\textit{Conclusion}}---We have measured magnetic-field-dependent Hall responses in YIG/Pt and YIG/Ta bilayers in which the opposite sign of the HSP signal and the thickness evolution of the MSP signal indicate contributions beyond the conventional even-in-$\theta_{\text{sh}}$ double spin Hall picture. Combining a magnetic-field-window separation [Fig.~\ref{fig:1b}] with the model of Eqs.~(\ref{dif})--(\ref{eq:odd_MSP}), we distinguish the OHE, HSP, and MSP contributions using sign conventions set by $R_H$, and thereby identify contributions odd in $\theta_{\text{sh}}$. The opposite signs of the Hall contributions in YIG/Ta and YIG/Pt are consistent with a component linear in the spin Hall angle, as described by a model incorporating interface-generated spin currents. Although this sign reversal alone does not exclude $G_i$-related interfacial exchange effects, the model comparison favors the odd-in-$\theta_{\text{sh}}$ single-SHE (IGSC+ISHE) interpretation. Fits based on $G_i$ require $G_i/G_r$ ratios well above the reported range, whereas the IGSC+ISHE scenario describes the data with $G_r$ fixed to literature values. More broadly, the combined analysis of HSP and MSP responses provides a route to disentangle bulk and interfacial spin-conversion pathways in MI/HM Hall magnetotransport, essential for both microscopic interpretation and the engineering of spintronic readout signals.

\textit{Acknowledgments}--- All authors from Mainz gratefully acknowledge funding support from the Deutsche Forschungsgemeinschaft (DFG) under the framework of the Collaborative Research Center TRR 173--268565370 \emph{Spin+X} (Projects B02, A01 and A12) and TRR 288--422213477 \emph{Elasto-Q-Mat} (Project A12), as well as from the European Union through the European Research Council (ERC) Synergy Grant No.~856538 (\emph{3DMAGiC}). We thank M.~Ossenschmidt, P.~Bougiatioti, S.~Soni, J.~Shan, and B.~J.~van~Wees for their contributions to the preliminary studies on the non-local anomalous Hall effect carried out at Bielefeld University and the University of Groningen, which were instrumental in enabling this research. V.A., E.M., and B.F. acknowledge support from the National Science Foundation under Grant ECCS-2236159.

The authors declare that they have no conflicts of interest.

\textit{Data availability}---The data supporting the findings of this study will be openly available.

\bibliography{main_bib}

\end{document}